\documentclass[pra,aps,twocolumn,showpacs,epsfig,superscriptaddress]{revtex4-1}
\usepackage{color,hyperref}
\usepackage{graphicx,amssymb,epsfig,epsf,amsmath}

\begin{document}

\title{Spectral fluctuation and $\frac{1}{f^{\alpha}}$ noise in the energy level statistics \\ 
of interacting trapped bosons} 

\title{Spectral fluctuation and $\frac{1}{f^{\alpha}}$ noise in the energy level statistics of interacting trapped bosons} 

\author{Kamalika Roy}
\affiliation{Department of Physics, Lady Brabourne College, P1/2 Surawardi Avenue, Kolkata 700017, India}

\author{Barnali Chakrabarti}

\affiliation{Department of Physics, Lady Brabourne College, P1/2 Surawardi Avenue, Kolkata 700017, India}

\affiliation{Instituto de Fisica, Universidade de S\~ao Paulo, CP 66318, 05315-970, S\~ao Paulo, SP Brazil}

\author{Anindya Biswas}
\thanks{corresponding author email: anindyabiswas@hri.res.in}
\affiliation{Department of Physics, University of Calcutta, 92 A.P.C. Road, 
Calcutta-700009, India}
\affiliation{Harish-Chandra Research Institute, Chhatnag Road, Jhunsi, Allahabad 211019, India}
\author{V. K. B. Kota}
\affiliation{Physical Research Laboratory, Navarangpura, Ahmedabad 380009, India}

\author{Sudip Kumar Haldar}
\affiliation{Department of Physics, Lady Brabourne College, P1/2 Surawardi Avenue, Kolkata 700017, India}

\begin{abstract} 
It has been recently shown numerically that the transition from integrability to chaos in quantum systems and the 
corresponding spectral fluctuations are characterized by $\frac{1}{f^{\alpha}}$ noise with $1\leq\alpha\leq 2$. 
The system of interacting trapped bosons is inhomogeneous and a complex system. The presence of external harmonic trap makes it more interesting as in the atomic trap the bosons occupy partly degenerate single-particle states. Earlier theoretical and experimental results show that at zero temperature the low-lying levels are of collective nature and high-lying excitations are of single particle nature. We observe that for few bosons, $P(s)$ distribution shows the Shnirelman peak which exhibits a large number of quasi-degenerate states. For large number of bosons the 
low-lying levels are strongly affected by the interatomic interaction and the corresponding level fluctuation shows a transition to Wigner with increase in particle number. It does 
not follow GOE (Gaussian Orthogonal Ensemble) Random Matrix predictions. For high-lying levels we observe the uncorrelated Poisson distribution. Thus it may be a very realistic system to prove that $\frac{1}{f^{\alpha}}$ noise is ubiquitous in nature.  
\end{abstract}
\maketitle
\section{Introduction}
\hspace*{.5cm}
Although there is no precise definition of quantum chaos, however it is closely related with the energy level 
fluctuation properties of a quantum system. Bohigas conjectured that level 
fluctuation of quantum system whose classical limit is chaotic, is described by the random matrix theory (RMT)~\cite{oboh}, whereas spectral fluctuation of classically integrable system obeys Poisson statistics~\cite{Sto}.
The concept of quantum chaos plays an important role in the understanding of the universal properties of 
the energy level spectrum of quantum system. However the complex natural systems are neither 
fully integrable, nor fully chaotic and attains special interest. RMT introduced by Wigner has been widely used in the description of complex spectrum of atomic nucleus, atoms and molecules~\cite{brody,vkbk,gom}. On the other hand, the bosonic ensembles in the dense limit may be ergodic with increase in the number of single particle states ~\cite{kota}. In last few years, interacting bosonic systems got special interest due to the experimental observation of Bose-Einstein condensation ~\cite{cor1, dvs, mha, cor2}. The presence of external harmonic trap makes it more interesting as stated by Asaga that in an atomic trap, bosonic atoms occupy partly degenerate single-particle states ~\cite{asaga}. Although it is argued that random-matrix approach should reveal the generic features of the spectrum however there is neither analytical treatment, nor systematic numerical calculations in this direction. The chaotic signature in the time evolution of Bose-Einstein condensation driven by the time-periodic harmonic or kicked pulses are observed ~\cite{pra79, pre73, pra81}. But energy level statistics of the experimentally dilute BEC has not been studied yet. In the earlier analysis of Bohigas in nuclear and atomic spectra the nearest-neighbor spacing distribution agrees very well with GOE (Gaussian Orthogonal Ensemble)~\cite{bohrev}. However for the interacting trapped bosons, it seems to contradict the usual expectation based on RMT. Very recently spectral properties of trapped 1D ultracold fermions in optical lattices are studied and the interplay of repulsive interaction with the external harmonic trap is observed~\cite{aya}. So it is also very much interesting to study the energy-level statistics of trapped bosons which are spatially inhomogeneous and we may expect new and rich physics. \\
\hspace*{.5cm}
Recently a different approach to characterize quantum chaos has been proposed based on the idea that the corresponding energy level sequence is analogous to the discrete time series. The level fluctuation is well characterized by the 
Fourier power spectrum and a power law behavior has been identified. This is conjectured that spectral fluctuations of chaotic quantum systems are characterized by 
$\frac{1}{f}$ noise whereas complete integrable systems exhibit $\frac{1}{f^{2}}$ noise ~\cite{arel,arel1,jmg,arel2}. The earlier studies in this 
direction involve quantum billiards, nonintegrable coupled quartic oscillator, kicked top, integrable spin chain~\cite{jmg,mss,graham,hs}. 
In this present work we study the system of $N$ interacting bosons at zero temperature in the presence of external trap. 
The choice of such system is important for various reasons. Firstly: it is inhomogeneous and complex system due to the presence of two 
energy scales. Interatomic interaction is characterized by $Na_{s}$, where $a_{s}$ is the $s$-wave scattering length and the external trap energy is characterized 
by $\hbar\omega$, where $\omega$ is the external trap frequency. From the earlier theoretical and experimental results it is an established fact that at zero temperature 
the low-lying collective excitations are strongly affected by the interatomic interaction when the high-lying excitations 
are of single particle nature ~\cite{fda1,graham,dsj,fda2}. The transition  from collective to single particle excitations makes us more curious 
to study the level fluctuation and to verify whether $\frac{1}{f^{\alpha}}$ noise is ubiquitous in nature. 
Secondly: the system directly manisfests the 
experimental Bose-Einstien condensation ~\cite{cor1,dvs,mha,cor2}. For the present calculation we consider the $N$-body bosonic system at zero temperature. There may be a very small effect of thermal cloud around the condensate even at zero temperature and the condensate is depleted due to the interaction~\cite{damping}. However for the present calculation we ignore that as the whole condensate is described by a single and fixed scattering length and the condensate is extremely dilute. Thus the effect of damping does not appear in our present calculation. However the effect of damping may be important when the interaction is tuned by external magnetic field. Thus the system in our present work is neither fully chaotic (for low-lying levels) nor fully 
integrable (for high-lying levels) due to the interplay of two-energy scales. At this point we should mention that Bohigas analysed thoroughly the nuclear shell-model and neutron resonance data for different nuclei. The nearest-neighbor spacing distribution of the nuclear data ensemble (NDE) agrees very well with the GOE prediction ~\cite{bohrev}. In the atomic spectra the levels with the same quantum numbers also show Wigner type spacing distribution. Thus in nuclear and atomic spectra, the regular features of low-lying part of the spectrum and chaotic features of high-lying collective levels are well established fact. However for the interacting trapped bosons, it seems to contradict the usual expectation based on RMT as for the experimental BEC, the low-lying excitations are collective where the interatomic interaction plays a crucial role and the high-lying levels are of single particle nature due to the dominating effect of external harmonic trap.\\

\hspace*{.5cm}
The paper is organized as follows. Sec.II deals with the methodology which contains the many-body technique to calculate the energy levels. Choice of interaction and the correlation function are also discussed in the same section. In the Sec.III we discuss several statistical tools and results. Sec.IV concludes the summary.\\ 

\section{Methodology}
\subsection{Many-body calculation with potential 
harmonic basis}
In order to calculate the energy levels of the condensate we solve the Schr\"odinger equation by our newly developed  correlated potential harmonic expansion method (CPHEM) with a short-range correlation function. CPHEM has already been established as a very successful technique for the study of dilute BEC ~\cite{tkd1, tkd2,tkd3}. In this method we keep all possible two-body correlation and also use a realistic interatomic interaction which is clearly an improvement over the mean-field Gross-Pitaevskii (GP) theory ~\cite{fda1,c6}. We briefly discuss the technique below.

We consider a system of $A=(N+1)$ identical bosons interacting via two-body potential $V(\vec{r}_{ij})$ = $V(\vec{r}_i-\vec{r}_j)$ and confined in an external harmonic potential of frequency $\omega$. The time-independent quantum many-body Schr\"odinger equation is given by
\begin{eqnarray}
\Big[-\frac{\hbar^2}{2m}\sum_{i=1}^{A} \nabla_{i}^{2} 
+ \sum_{i=1}^{A} V_{trap}(\vec{r}_{i})& 
+\displaystyle{\sum_{i,j>i}^{A}} V(\vec{r}_{i}-\vec{r}_{j})\nonumber\\
-E\Big]\Psi(\vec{r}_{1},...,\vec{r}_{A})=0\hspace*{.1cm};
\end{eqnarray}
where $m$ is the mass of the each boson and $E$ is the energy of the condensate. After eliminating the center of mass motion by using the standard Jacobi vectors~\cite{fabre,fab2,fab3}, defined by 
\begin{equation}
\vec{\zeta}_{i}=\sqrt{\frac{2i}{i+1}}(\vec{r}_{i+1}-
\frac{1}{i}\sum_{j=1}^{i} \vec{r}_j) \hspace*{.5cm}
 (i=1,...N),
\end{equation}
we obtain the relative motion of $N$-body system as 
\begin{eqnarray}
\Big[-\frac{\hbar^{2}}{m}\sum_{i=1}^{N} 
\nabla_{\zeta_{i}}^{2}+V_{trap}& + &V_{int}
(\vec{\zeta}_{1}, ..., \vec{\zeta}_{N})\nonumber\\
-E_{R}\Big]\Psi(\vec{\zeta}_{1}, ..., \vec{\zeta}_{N})& = & 
0\hspace*{.1cm}, 
\end{eqnarray}
$V_{trap}$ is the effective external trapping potential, $V_{int}$ is the sum of all pair-wise interactions expressed in terms of the Jacobi vectors and  $E_R$ is the relative energy of the system {\it i.e.} $E=E_R+\frac{3}{2}\hbar\omega$.

Now it is to be noted that Hyperspherical harmonic expansion method (HHEM) is an {\it ab-initio} tool to solve the many-body Schr\"odinger equation where the total wave function is expanded in the complete set of hyperspherical basis~\cite{fab2}. Although HHEM is a complete many-body approach and includes all possible correlations, it can not be applied to a typical BEC containing few thousands to few millions of bosons. Due to the large degeneracy of the HH basis, HHEM is restricted only to three-particle systems~\cite{fab2,esry}. Since the typical experimental BEC is designed to be very dilute and the probability of three and higher-body collisions is negligible, we can safely ignore the effect of three and higher-body correlations. Therefore only two-body correlation and pairwise interaction among the bosons is important. It allows us to decompose the total wave function $\Psi$ into two-body Faddeev component for the interacting $(ij)$ pair as 
\begin{equation}
\Psi=\sum_{i,j>i}^{A}\phi_{ij}(\vec{r}_{ij},r)\hspace*{.1cm}\cdot
\end{equation}
It is worth to note that $\phi_{ij}$ is a function of two-body separation ($\vec{r}_{ij}$) and the global hyperradius $r$ is given by, $r = \sqrt{\sum_{i=1}^{N}\zeta_{i}^{2}}$. Thus the effect of two-body correlation comes through the two-body interaction in the expansion basis. $\phi_{ij}$ is symmetric under $P_{ij}$ for bosons and satisfy the Faddeev equation
\begin{equation}
\left[T+V_{trap}-E_R\right]\phi_{ij}
=-V(\vec{r}_{ij})\sum_{k,l>k}^{A}\phi_{kl}
\end{equation}
where $T = -\frac{\hbar^2}{m} \displaystyle{\sum_{i=1}^{N}} \nabla_{\zeta_{i}}^{2}$ is the total kinetic energy. Operating $\sum_{i,j>i}$ on both sides of equation (5), we get back the original Schr\"odinger equation. In this approach, we assume that when ($ij$) pair interacts, the rest of the bosons are inert spectators. Thus the total hyperangular momentum 
quantum number as also the orbital angular momentum of the whole system is contributed by the interacting pair only. Next  we expand $\phi_{ij}$ in the subset of hyperspherical harmonics (HH) necessary for the expansion of $V(\vec{r}_{ij})$.
\begin{equation}
\phi_{ij}(\vec{r}_{ij},r)
=r^{-(\frac{3N-1}{2})}\sum_{K}{\mathcal P}_{2K+l}^{lm}
(\Omega_{N}^{ij})u_{K}^{l}(r) \hspace*{.1cm}\cdot
\end{equation}
$\Omega_N^{ij}$ denotes the full set of hyperangles in the $3N$-dimensional 
space corresponding to the $(ij)$th  interacting pair and 
${\mathcal P}_{2K+l}^{lm}(\Omega_N^{ij})$ is called the PH basis. It 
has an analytic expression:
\begin{equation}
{\mathcal P}_{2K+l}^{l,m} (\Omega_{N}^{(ij)}) =
Y_{lm}(\omega_{ij})\hspace*{.1cm} 
^{(N)}P_{2K+l}^{l,0}(\phi) {\mathcal Y}_{0}(D-3) ;\hspace*{.5cm}D=3N ,
\end{equation}
${\mathcal Y}_{0}(D-3)$ is the HH of order zero in 
the $(3N-3)$ dimensional space spanned by $\{\vec{\zeta}_{1}, ...,
\vec{\zeta}_{N-1}\}$ Jacobi vectors; $\phi$ is the hyperangle given by
$r_{ij}$ = $r\hspace*{0.1cm} cos\phi$. For the remaining $(N-1)$
 noninteracting bosons we define hyperradius as
\begin{eqnarray}
 \rho_{ij}& = &\sqrt{\sum_{K=1}^{N-1}\zeta_{K}^{2}}\nonumber\\
          &= & r \sin\phi \hspace*{.01 cm}\cdot
\end{eqnarray}
such that $r^2=r_{ij}^2+\rho_{ij}^2$ and $r$ represents the global 
hyperradius of the condensate. The set of $(3N-1)$ quantum 
numbers of HH is now reduced to {\it only} $3$ as for the $(N-1)$ 
non-interacting pair
\begin{eqnarray}
l_{1} = l_{2} = ...=l_{N-1}=0,   & \\
m_{1} = m_{2}=...=m_{N-1}=0,  &   \\
n_{2} = n_{3}=...n_{N-1} = 0, & 
\end{eqnarray}
and for the interacting pair $l_{N} = l$, $m_{N} = m$ and  $n_{N} = K$.
Thus the $3N$ dimensional Schr\"odinger equation reduces effectively
to a four dimensional equation with the relevant set of quantum 
numbers: hyperradius $r$, orbital angular momentum quantum number $l$,
azimuthal quantum number $m$ and grand orbital quantum number $2K+l$
 for any $N$.
Substituting Eq(6) into Eq.(5) and projecting on a particular PH, a set of 
coupled differential equation (CDE) for the partial wave $u_{K}^{l}(r)$
is obtained
\begin{equation}
\begin{array}{cl}
&\Big[-\frac{\hbar^{2}}{m} \frac{d^{2}}{dr^{2}} +
V_{trap}(r) + \frac{\hbar^{2}}{mr^{2}}
\{ {\cal L}({\cal L}+1) \\
&+ 4K(K+\alpha+\beta+1)\}-E_R\Big]U_{Kl}(r)\\
+&\displaystyle{\sum_{K^{\prime}}}f_{Kl}V_{KK^{\prime}}(r)
f_{K^{\prime}l}
U_{K^{\prime}l}(r) = 0
\hspace*{.1cm},
\end{array}
\end{equation}\\
where ${\mathcal L}=l+\frac{3A-6}{2}$, $U_{Kl}=f_{Kl}u_{K}^{l}(r)$, 
$\alpha=\frac{3A-8}{2}$ and $\beta=l+1/2$.\\
$f_{Kl}$ is a constant and represents the overlap of the PH for
interacting partition with the sum of PHs corresponding  to all 
partitions~\cite{fab3}.
The potential matrix element $V_{KK^{\prime}}(r)$ is given by
\begin{equation}
V_{KK^{\prime}}(r) =  
\int P_{2K+l}^{lm^*}(\Omega_{N}^{ij}) 
V\left(r_{ij}\right)
P_{2K^{\prime}+1}^{lm}(\Omega_{N}^{ij}) d\Omega_{N}^{ij} 
\hspace*{.1cm}\cdot
\end{equation}\\

\subsection{Choice of interaction and introduction of additional short range correlation}

In the mean-field GP equation the two-body interaction is taken as the contact $\delta$ potential, the interaction strength being proportional to the $s$-wave scattering length $a_{s}$. A positive value of $a_{s}$ gives a repulsive condensate and a negative value of $a_{s}$ gives an attractive condensate. But the contact interaction completely disregards the detailed structure. However a realistic interatomic interaction, like the van der Waal potential, is always associated with an attractive 
$-\frac{C_6}{{r_{ij}}^6}$ tail at large separation and a strong repulsion at short separation. Depending on the nature of these two parts, $a_{s}$ can be either positive or negative. In our earlier calculations \cite{tkd2008} we have already observed the effect of shape-dependent interatomic interaction in the many-body calculation. So for our present calculation we choose the van  der Waal potential with a hard core repulsion of radius $r_c$, {\it viz}, $V(r_{ij})$ = $ \infty $ for $r_{ij}$ $\leq$ $r_c$ and $-\frac{C_6}{{r_{ij}}^6}$ for $r_{ij}$ $>$ $r_c$. 
The value of $C_6$ is fixed for a given system and for $^{87}$Rb atoms $C_6=6.4898\times 10^{-11}$ o.u. ~\cite{c6}. Throughout our calculation we choose $a_{ho}$ = $\sqrt{\frac{\hbar}{m\omega}}$ as the unit of length (o.u.) and energy is also expressed in the unit of oscillator energy ($\hbar\omega$). For a given two-body interaction  $a_{s}$ can be obtained from the solution of two-body equation with zero energy. 
\begin{equation}
-\frac{\hbar^2}{m}\frac{1}{r_{ij}^2}\frac{d}{dr_{ij}}\left(r_{ij}^2
\frac{d\eta(r_{ij})}{dr_{ij}}\right)+V(r_{ij})\eta(r_{ij})=0
\hspace*{.1cm}\cdot
\end{equation} 
 The solution of the two-body equation shows that the value of $a_{s}$ changes from negative to positive and thus passing through an infinite discontinuity as $r_c$ decreases [Fig. 1]. At each discontinuity one extra node appears in the two-body wave function which corresponds to one extra two-body bound state. With a tiny increase in $r_c$, across the infinite discontinuity  $a_{s}$ changes drastically from a very large positive value to a large negative value and the properties of the condensate changes drastically ~\cite{c6}. In the GP equation one uses $a_{s}$ directly without any such detailed knowledge of actual interatomic potential. For the present calculation we choose $a_{s}$=0.00433 o.u. which mimics the JILA experiment with $^{87}$Rb atoms~\cite{mha}.
 The corresponding value of $r_c$ is 1.121$\times10^{-3}$ o.u. which causes one node in the two-body wave function. The normalization constant is chosen to make the wave function positive at large $r_{ij}$.\\
\hspace*{.5cm}
 In the experimental BEC, the Bose gas is extremely dilute, the average interparticle separation is much larger than the range of the two-body interaction. This is required to prevent 
the three-body collision and formation of molecules. Thus the pair of particles with practically zero kinetic energy do not come closer than $a_{s}$. Whereas the zero$^{th}$ order PH is a constant \cite{fabre} and will give a large probability even for $r_{ij}$$\rightarrow0$, it causes very slow convergence in the PH basis [Eq.(6)]. To compensate this we additionally include a short range correlation function $\eta(r_{ij})$ in the PH expansion. As the fundamental assumption in our method is to consider only ($ij$) pair interaction when the remaining particles are simply inert spectators, the correlation function is obtained as the zero-energy solution of the two-body equation [Eq.(14)]. The correlation function quickly attains asymptotic form $(1-\frac{a_s}{r_{ij}})$ for large $r_{ij}$. We replace Eq(6) by  

\begin{equation}
\phi_{ij}(\vec{r}_{ij},r)
=r^{-(\frac{3N-1}{2})}\sum_{K}{\mathcal P}_{2K+l}^{lm}
(\Omega_{N}^{ij})u_{K}^{l}(r) \eta(r_{ij}) \hspace*{.1cm}\cdot
\end{equation}
The correlated PH (CPH) basis becomes
\begin{equation}
[{\mathcal P}_{2K+l}^{l,m} (\Omega_{N}^{(ij)})]_{correlated} =
 {\mathcal P}_{2K+l}^{l,m} (\Omega_{N}^{(ij)}) \eta(r_{ij}) ,
\end{equation}
The correlated potential matrix $V_{KK^{\prime}}(r)$ is now given by
\begin{equation}
\begin{array}{cl}
&V_{KK^{\prime}}(r) =(h_{K}^{\alpha\beta} h_{K^{\prime}}^
{\alpha\beta})^{-\frac{1}{2}}\times \\
&\int_{-1}^{+1} \{P_{K}^{\alpha\beta}(z) 
V\left(r\sqrt{\frac{1+z}{2}}\right)
P_{K^{\prime}}^{\alpha \beta}(z)\eta\left(r\sqrt{\frac{1+z}{2}}\right)
W_{l}(z)\} dz \hspace*{.1cm}\cdot
\end{array}
\end{equation}
Here $P_{K}^{\alpha\beta}(z)$ is the Jacobi polynomial, and its 
norm and 
weight function are $h_{K}^{\alpha\beta}$ and $W_{l}(z)$   
respectively~\cite{abraham}.\\
\hspace*{.5cm}
One may note that the inclusion of $\eta(r_{ij})$ makes the PH basis 
non-orthogonal. One may surely use the standard procedure for handling 
non-orthogonal basis. However in the present calculation we have 
checked that $\eta(r_{ij})$ differs from a constant value only by small 
amount and the overlap $\Big< {\mathcal P}_{2K+l}^{l,m} (\Omega_{N}^{(ij)})|{\mathcal P}_{2K+l}^{l,m} (\Omega_{N}^{(kl)})\eta(r_{kl})\Big>$ is quite small. Thus we get back the Eq(12) 
approximately when the correlated potential matrix is calculated by Eq(17). \\
\hspace*{.5cm}
Finally the coupled differential equation (CDE), 
Eq.~(12), is solved by the hyperspherical adiabatic approximation 
(HAA)~\cite{tkd5}. 
In HAA, one assumes that the hyperradial motion is slow compared 
to the 
hyperangular motion. Hence the latter is separated adiabatically 
and solved 
for a particular value of $r$, by diagonalizing the potential matrix 
together with the diagonal hypercentrifugal repulsion in Eq.~(12). 
The lowest eigenvalue, $\omega_0(r)$ is the effective potential 
for the hyperradial motion
 and in this effective potential 
the entire 
condensate moves as a single entity. Thus in HAA, the approximate 
solution 
(the energy and wave function) of the condensate is obtained by solving a 
single uncoupled differential equation
\begin{equation}
\left[-\frac{\hbar^{2}}{m}\frac{d^{2}}{dr^{2}}+\omega_{0}(r)-E_{R}
\right]\zeta_{0}(r)=0\hspace*{.1cm},
\end{equation}
subject to appropriate boundary conditions on $\zeta_{0}(r)$. 
The function $\zeta_{0}(r)$ is the collective wave function of 
the condensate in the hyperradial space. 
The lowest lying state in the effective potential $\omega_{0}(r)$ 
corresponds to the ground state of the condensate. The total energy 
of the condensate is obtained by adding the energy of the center of 
mass motion $(\frac{3}{2}\hbar\omega)$ to $E_{R}$.\\

Thus by employing the CPHEM and HAA we reduce the multi-dimensional problem into an effective one-dimensional problem in hyperradial space and the effective potential $\omega_{0}(r)$ provides both the qualitative and quantitative description of the system. As in our many-body picture, the collective motion of the condensate is characterised by the effective potential, the excited states in this potential are the states with ${n}^{th}$ radial excitation and ${l}^{th}$ surface mode and are generally denoted by $E_{nl}$. Thus $E_{00}$ corresponds to the ground state and $l\neq 0$ corresponds to several surface modes. For $l>0$, we calculate the potential matrix from the diagonal hypercentrifugal term. We have checked that the contribution coming from the off-diagonal matrix element is very very small and we disregard these matrix elements as they make the computation very slow. The calculation of low-lying collective modes are in good agreement with the experimental results and other calculations~\cite{pan, abi1}. For energy much larger than the chemical potential ($\mu$) we observe that the states are separated at energy close to harmonic oscillator energies ($\sim \hbar\omega$). This transition from the low-energy collective modes to high-lying single particle excitation are further used for the statistical calculations.\\

\section{Results}
\hspace*{.5cm}
The integrated level density $N(E)$ has two parts. One is the smooth part ($\bar{N}(E)$) and a fluctuating part ($\tilde{N}(E)$). To compare the
fluctuation of different systems or different parts of the same system, the smooth part is removed by the unfolding 
procedure. Unfolding maps the energy levels $E_{i}$ to $\epsilon_{i}$ with the unit mean level density. For the present 
analysis the many-body level density is approximated by a polynomial and unfolding is done by $7^{th}$ order polynomial. 
We unfold each spectrum separately for a specific value of $l$ and form an ensemble having the same symmetry. Then the nearest neighbor spacing is calculated as $s_{i}=\epsilon_{i+1}-\epsilon_{i}$, 
$i=1,2,...n$. For the further study of correlation and level-repulsion between energy levels we utilize the established analogy between the energy spectrum and discrete time series~\cite{arel,arel1,jmg,efa}. The energy spectrum is considered as a discrete signal 
and the fluctuations of the excitation energy as discrete time series. The $\delta_{n}$ statistics has been 
used in RMT to study how the consecutive level spacings are correlated. It is defined as 
\begin{equation}
\delta_{n} = \sum_{i=1}^{n}(s_{i}-<s>) = \epsilon_{n+1}-\epsilon_{1}-n
\end{equation}
As the average value of $s_{i}$ is $<s>=1$, $\delta_{n}$ represents the deviation of $(n+1)^{th}$ level from the 
mean value i.e. the fluctuation of $(n+1)^{th}$ excited state. 
It is also closely related to the level density fluctuations and one can write $\delta_{n}$=$-\tilde{N}(E_{n+1})$ if the ground state energy is shifted appropriately~\cite{efa}. Thus it represents the accumulated level density fluctuation at $E$=$E_{n+1}$. $\delta_{n}$ is similar to the time series and $n$ 
represents the discrete time ~\cite{arel1,jmg,arel2,efa}. The power spectrum is then defined as the square modulus of the Fourier transform as 
\begin{equation}
P_{k}^{\delta} = | \frac{1}{\sqrt{M}} \sum_{n} \delta_{n} exp(-\frac{2\pi ikn}{M})|^{2}
\end{equation}
where $k=1,2,....n$ and $f=\frac{2\pi k}{M}$ represents the frequency and $M$ is the size of the series ~\cite{jmg}. Therefore, the statistical behavior of level fluctuation can be established by $<P_{k}^{\delta}>$ statistic which measures both short and long range correlation. It is verified that the power laws ${P}_{k}^{\delta}$ $\propto$$\frac{1}{k^{\alpha}}$ both for fully chaotic and integrable systems~\cite{arel,arel1,jmg,arel2}. But depending on the level correlation in the chosen system $\alpha$ scales smoothly from 1 (chaotic system) to 2 (for uncorrelated and integrable system)~\cite{arel2,arel,arel1,jmg}. However in the integrable spin chains of Halden-Shastry type, the spectral fluctuations exhibit $\frac{1}{f^4}$ noise rather than the expected $\frac{1}{f^2}$ noise~\cite{hs}.\\

\hspace{.5cm}
In Fig. 2 we display the energy level fluctuations for different number of energy levels for 5000 bosons in the trap. For the low-lying 
levels we expect level correlation. As the low-lying levels are highly affected by the interatomic interaction, 
the energy spectra shows level repulsion and strong spectral rigidity. This is reflected in the Fig. 2(a) which looks like 
the antipersistent time series for the lowest 500 levels. The $\delta_{n}$ statistics for the low levels is very close to the GOE spectra which indicates high level correlation due to the interatomic interaction. For the intermediate levels, the effect of interatomic interaction gradually decreases 
and the external trap starts to dominate. Thus the system is expected to show a mixed and complex statistics. When a 
part of the levels are correlated due to interatomic interaction and two-body correlation, the other part do not 
repel each other and uncorrelated. It is similar to the classical mixed system, where a part of 
phase space is completely regular with the other part chaotic. Thus the Fig. 2(b) shows that $\delta_{n}$ is neither 
persistent nor antipersistent. For much higher levels [Fig. 2(c)], the energy levels are uncorrelated due to the dominating 
effect of the external harmonic trap. The system is close to integrable and $\delta_{n}$ looks like a persistent series of Poisson spectra. 
To characterise long-range correlation in Fig. 2(d)-2(f), we plot the average values of the power spectrum 
$<P_{k}^{\delta}>$ for the same number of levels as reported in Fig. 2(a)-2(c). It shows that the power spectrum follows the 
scaling law $<P_{k}^{\delta}>$ $\simeq$ $\frac{1}{k^{\alpha}}$. The value of $\alpha$ is presented in Fig. 2(d)-2(f) for different 
number of levels. For low-lying correlated levels $\alpha$ = 1.31, for intermediate levels $\alpha$ = 1.72 and for 
high-lying levels $\alpha$ = 1.99. Thus $\alpha$ not only measures the 
chaoticity of the system but it measures the degree of integrability for complex systems.
At this point we should mention that in a nice attempt the momentum distribution and temporal power spectra of nonzero temperature Bose-Einstein condensate are calculated using the Gross-Pitaevskii equation~\cite{dnoisebec}. The temporal power spectra also shows $\frac{1}{f^{\alpha}}$ form where $\alpha=2-\frac{D}{2}$ ($D$ is the dimension of space) ~\cite{dnoisebec, dnoisetemp}.
Next to compare the result with the most popular and well known statistics, we calculate the nearest neighbor spacing distribution $P(s)$ and plot in Fig.~3 and in Fig.~6. In an earlier attempt in this direction we have reported some preliminary results on level-spacing distribution $P(s)$ ~\cite{abips}. We have shown that due to interatomic correlation the lower levels are strongly affected by the interaction, however the higher levels are uncorrelated. But our earlier results do not prove the Asaga's statement which says that in an atomic trap, bosonic atoms are in partly degenerate single particle states. This makes us very curious to study in details how the small interatomic interaction will act as a perturbation and lift the degeneracy. This needs further numerical analysis for varying number of bosons and with increase in number of levels. In Fig.~3 we plot the $P(s)$ distribution for the lowest 100 levels for different number of bosons. For $N=3$ with $a_{s}$ = 2.09
$\times$ $10^{-4}$ o.u., the effective interaction $Na_{s}$ is 8.7 $\times$ $10^{-4}$ o.u. The system is very close to integrable as the effect of such small interaction is masked due to the effect of external harmonic trap. At zero temperature the interaction energy for $N=3$ is almost negligible compared to the trap energy. Thus the small interaction acts as a very small perturbation and the exact degeneracy in the external $3D$ harmonic trap is lifted and it results to the existence of large quasi-degenerate states. $P(s)$ distribution exhibits $\delta$-type peak called as Shnirelman peak. In the year 1993, Shnirelman showed that for systems with time reversal symmetry should exhibit such a $\delta$-function peak near $s=0$ in the $P(s)$ distribution. It is known as the Shnirelman peak. This peak appears due to the presence of symmetry and separating levels by symmetry, one will get back Poisson distribution. This indicates the presence of bulk quasi-degenerate states in the level spacing distribution. In the first verification of Shnirelman theorem, Chirikov and Shepelyansky studied the kicked rotator on a torus with time-reversal symmetry ~\cite{prl74}. Later the theorem is verified in a more real physical quantum system.
 The Calogero-like three-body problem was studied where the hidden continuous symmetry was broken by adding a three-body interaction term~\cite{pre2002}. With further increase in $N$ gradually in the trap, the lower levels show level-repulsion and the system smoothly changes to close to integrability to nonintegrability. The corresponding $P(s)$ distribution smoothly changes to Wigner like distribution with increase in $N$. Due to strong interatomic interaction the system becomes more correlated and show level-repulsion. \\
\hspace*{.5cm}
    In  our present problem of trapped, interacting bosons, the exact degeneracy comes from the external harmonic trap. However due to the weak interatomic interaction, the effect of exact degeneracy is gradually lifted and it results to the quasi-degeneracy when the number of bosons in the trap is quite small. For better resolution of the Shnirelman peak appeared in Fig.~3 (with $N$=3) we plotted the same in Fig.~4 in finer details. A huge peak in the first bin of the histograms clearly demonstrates the existence of global quasi-degeneracy in accordance with the Shnirelman theorem. In the top left-most panel in Fig.~4, we observe the peak has a finite width which is further associated with the Poissonian tail. This peak contains important information about the structure of the quantum system. The resolution of the peak is further plotted in Fig.~5 where we present the integral level-spacing distribution $I(s)=NP(s)$, normalized to unity. It has two separate regions. The rightmost steep-increase of $I(s)$ corresponds to the Poissonian tail of Fig.~4. The leftmost part is more interesting. It shows the linear dependence between $I$ and $ln (s)$, which represents the structure of the Shnirelman peak.\\
\hspace*{.5cm}
    The results for higher levels close to 4000 levels and for the same set of $N$ values as reported in Fig.~3, are plotted in Fig. 6.  For $N=3$ , $P(s)$ distribution again shows the sharp peak as expected. As the high lying excitations are of single particle nature, the energy levels are now uncorrelated and the corresponding $P(s)$ distribution shows Poisson type fluctuation with increase in number of bosons. It confirms that for higher levels the system again becomes close to integrable as the effect of external trap strongly dominates. The observation is in correlation with the earlier observation of $\delta_{n}$ statistics and power spectrum. $P(s)$ measures the short-range correlation. The $\Delta_{3}$ statistic is usually used to investigate the long-range 
correlation. It gives the statistical measure of the rigidity of finite spectral level sequence. For a given energy interval $L$, it is determined by the 
least square deviation of the staircase from the best straight line fits it. In Fig. 7 we plot the spectral average $<\Delta_{3}(L)>$ for 
different energy levels. For higher energy levels $<\Delta_{3}(L)>$ bends to Poisson whereas for low-lying collective 
levels it close to GOE prediction.

\section{Conclusions}
\hspace*{.5cm}
In summary, we have shown that the analogy between quantum energy spectra and time series is an efficient and powerful way to characterize quantum level fluctuation. Although the statistical behaviour of level fluctuation and the corresponding power spectrum are understood for fully chaotic and completely integrable systems, the behaviors of power spectrum in the mixed regime between integrability and chaos is interesting. Interacting trapped bosons is a very complex system and due to the existence of two energy scales it nicely describes chaos to order transition with increase in number of energy levels. Our observation of Shnirelman peak strongly proves the earlier statement of Asaga ~\cite{asaga}. Our results nicely demonstrate how the degenerate single particle states of the pure harmonic trap are lifted gradually by increasing the effective interatomic interaction. Our findings are quite different from the results seen in atomic nuclei, atoms and molecules~\cite{bohrev}. Interacting trapped bosons is a very special and very complex system where the low-lying collective excitations are strongly influenced by interatomic interaction and shows level-repulsion. It is also spatially inhomogeneous and the high-lying levels are of single particle nature and have regular features . For the dilute interacting Bose gas, it is also possible to calculate a large number of energy levels with high statistical precision. They can also be measured experimentally.  The corresponding level fluctuation shows a transition from close to Wigner to Poisson with increase in energy levels showing  it does 
not follow GOE predictions and we need a modified GOE which combines uniform, GOE and Poisson ~\cite{tag}. We observe the existence of $\frac{1}{f^{\alpha}}$ power law in the energy spectrum. The parameter $\alpha$ measures the fluctuation properties of the quantum system through the power spectrum. As the interacting trapped bosons are interesting  in the connection of recent experiments of BEC, our system is generic and it confirms that $\frac{1}{f^{\alpha}}$ noise is ubiquitous in nature. However some open questions {\it viz.} how the  spectral distribution will change with the attractive interactions to study the dynamical behavior of energy spectrum, still remain . \\

\hspace*{.5cm}
The work is supported by D.A.E. (Grant no. 2009/37/23/BRNS/1903). AB acknowledges CSIR, India for a senior research fellowship (sanction no. 09/028(0773)-2010-EMR-1). 
S.K.H. acknowledges CSIR for junior research fellowship.\\

\hspace*{1cm}

\vskip 0.5cm

\begin{figure}
  \begin{center}
    \begin{tabular}{cc}  
    \resizebox{75mm}{!}{\includegraphics[angle=-90]{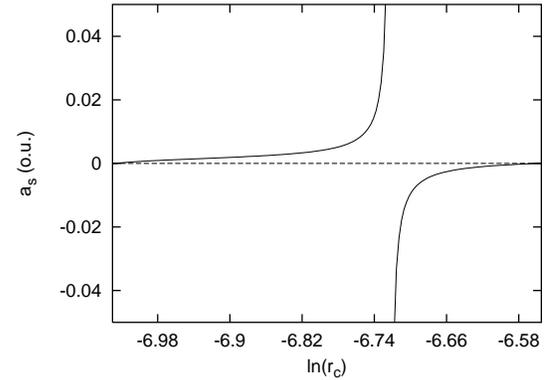}} \\	 
    \end{tabular}
  \end{center}
\caption[angle=0]{plot of scattering length ($a_{s}$) against $ln(r_c)$}
\end{figure}

\begin{figure}
  \begin{center}
    \begin{tabular}{cc}
      \resizebox{40mm}{!}{\includegraphics[angle=0]{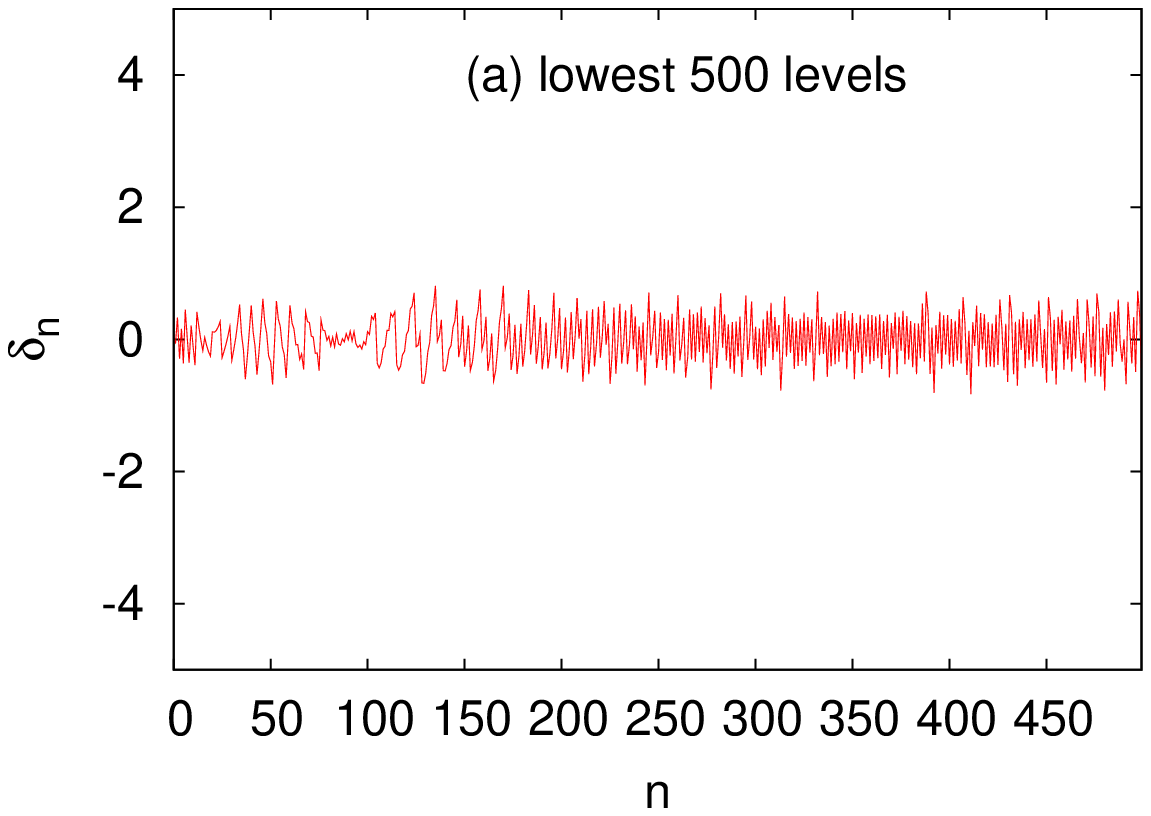}} &
      \resizebox{40mm}{!}{\includegraphics[angle=0]{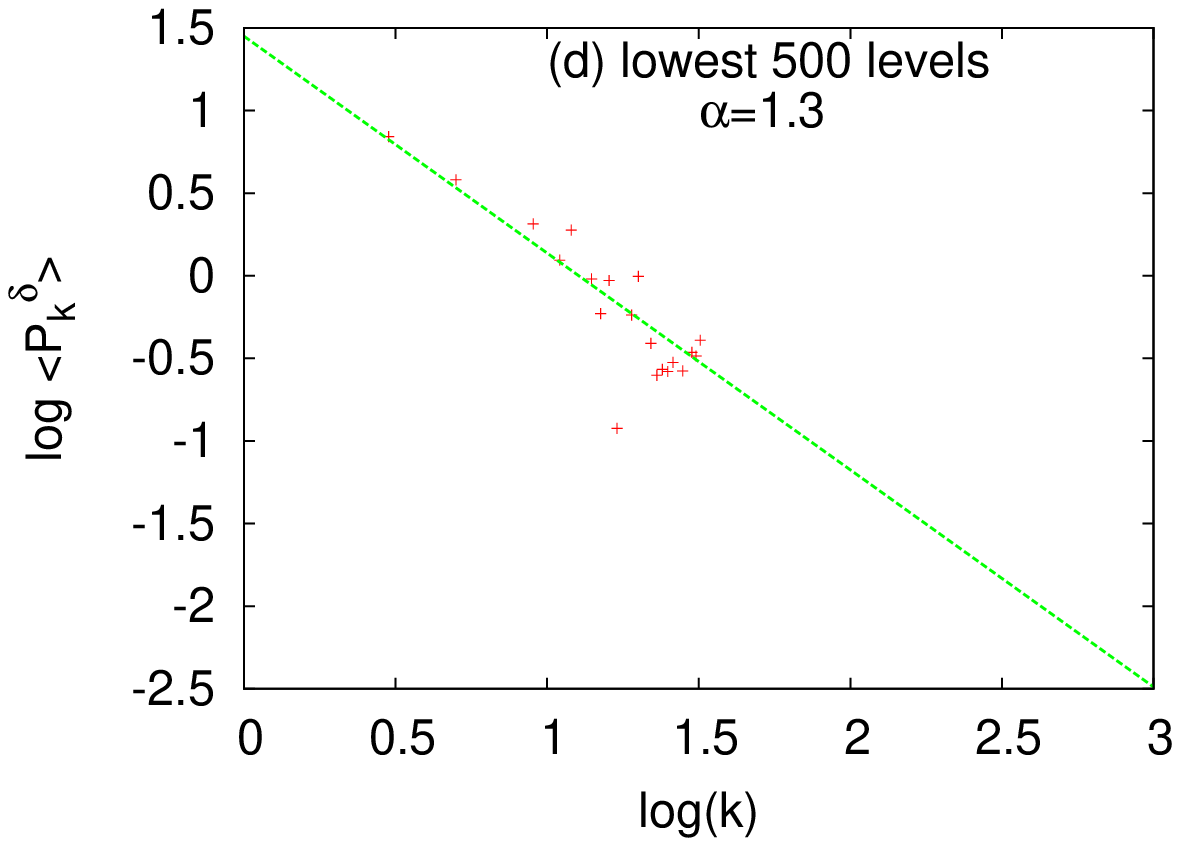}} \\
	 
      \resizebox{40mm}{!}{\includegraphics[angle=0]{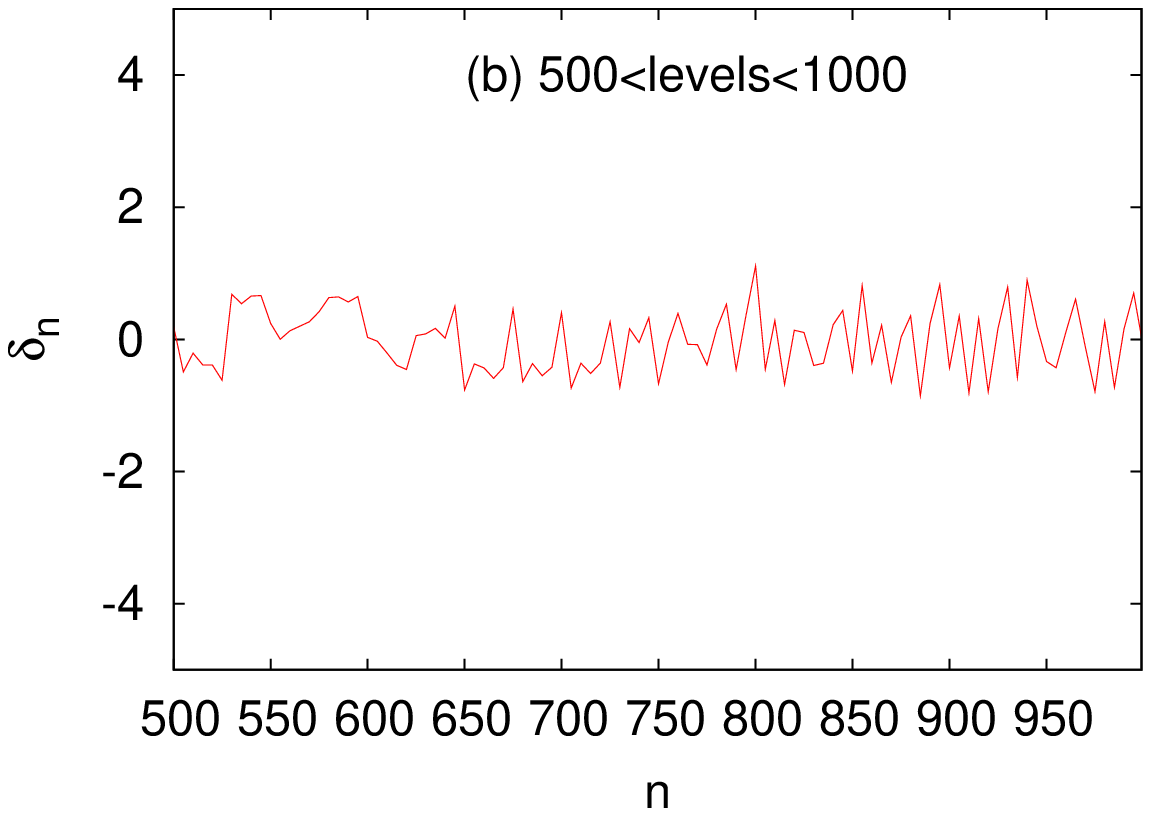}} &
      \resizebox{40mm}{!}{\includegraphics[angle=0]{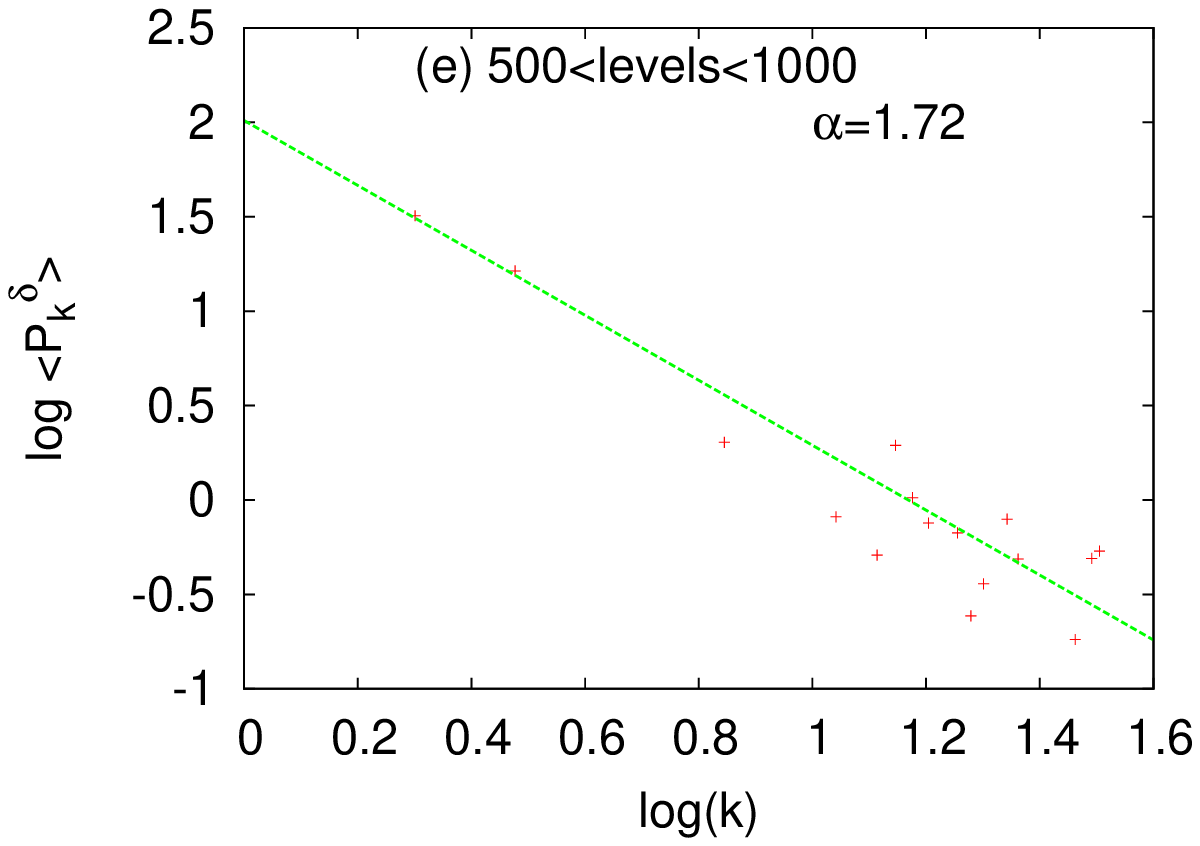}} \\
	
      \resizebox{40mm}{!}{\includegraphics[angle=0]{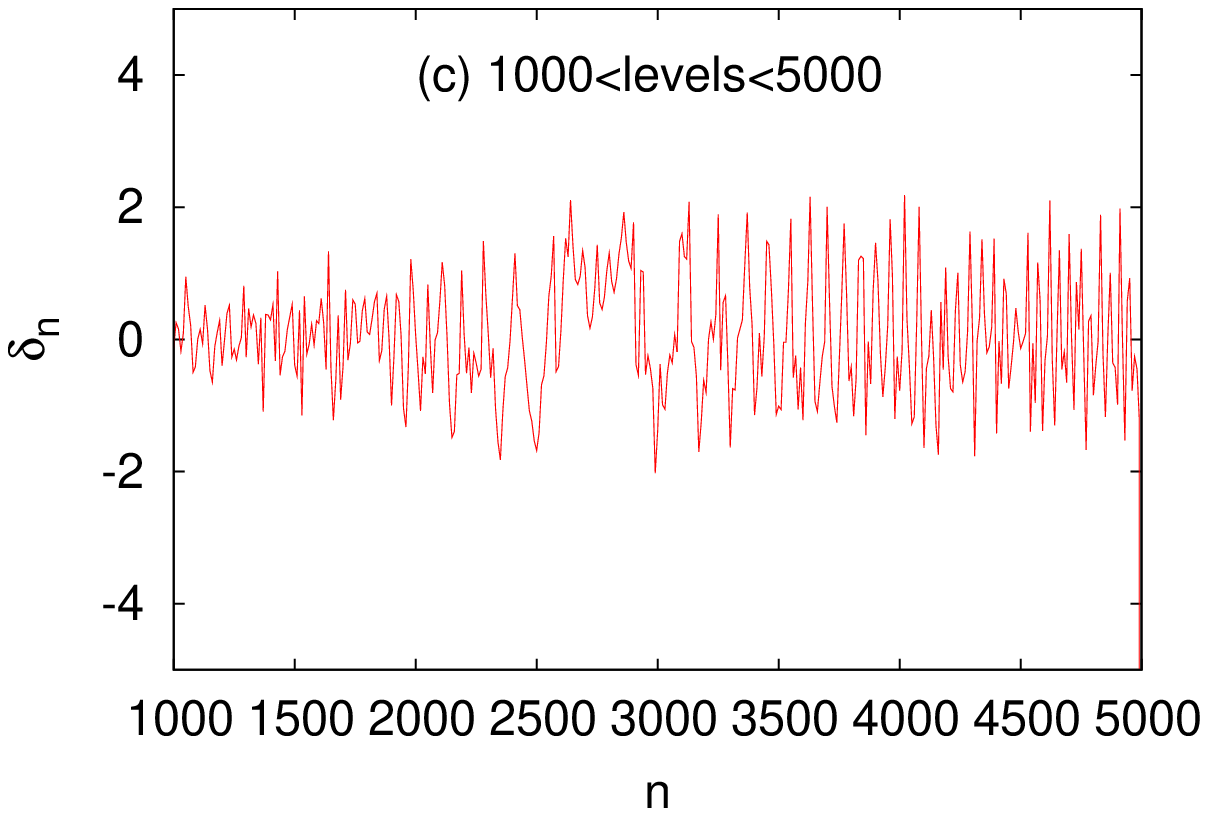}} &
      \resizebox{40mm}{!}{\includegraphics[angle=0]{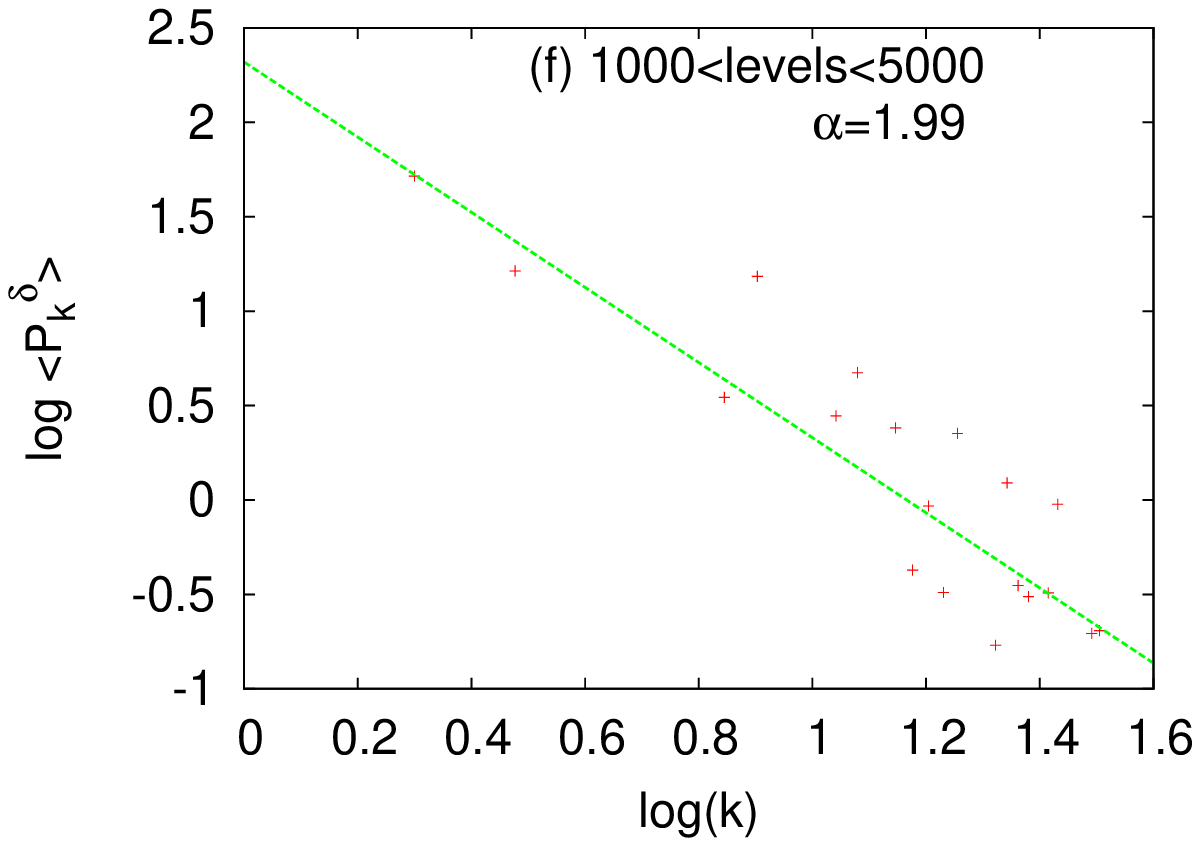}} \\
    \end{tabular}
  \end{center}
\caption[angle=90]{(color online) Fig.~1(a)-(c): Plot of fluctuation $\delta_{n}$ statistics for different energy levels. Fig.~1(d)-(f): Plot of average power spectrum.}
\end{figure}

\vskip 0.5cm
\begin{figure}
  \begin{center}
    \begin{tabular}{cc}
      \resizebox{40mm}{!}{\includegraphics[angle=0]{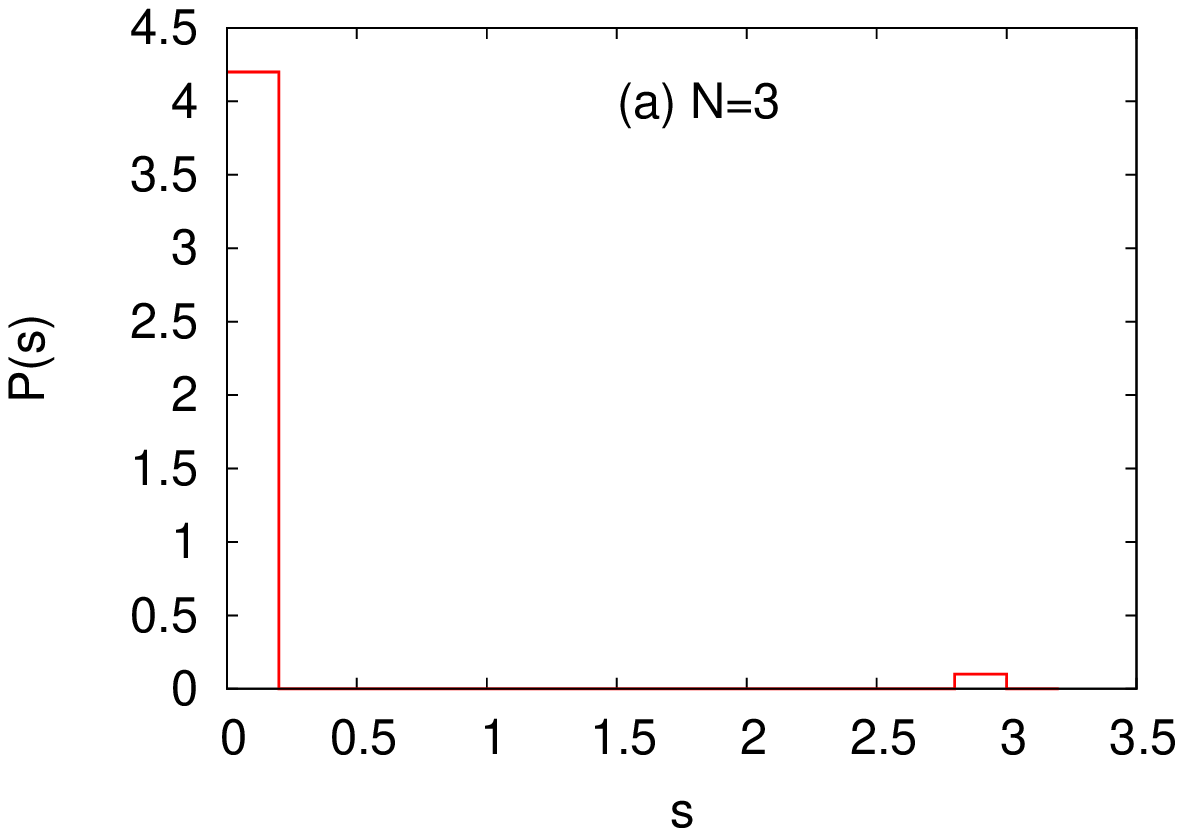}} &
       \resizebox{40mm}{!}{\includegraphics[angle=0]{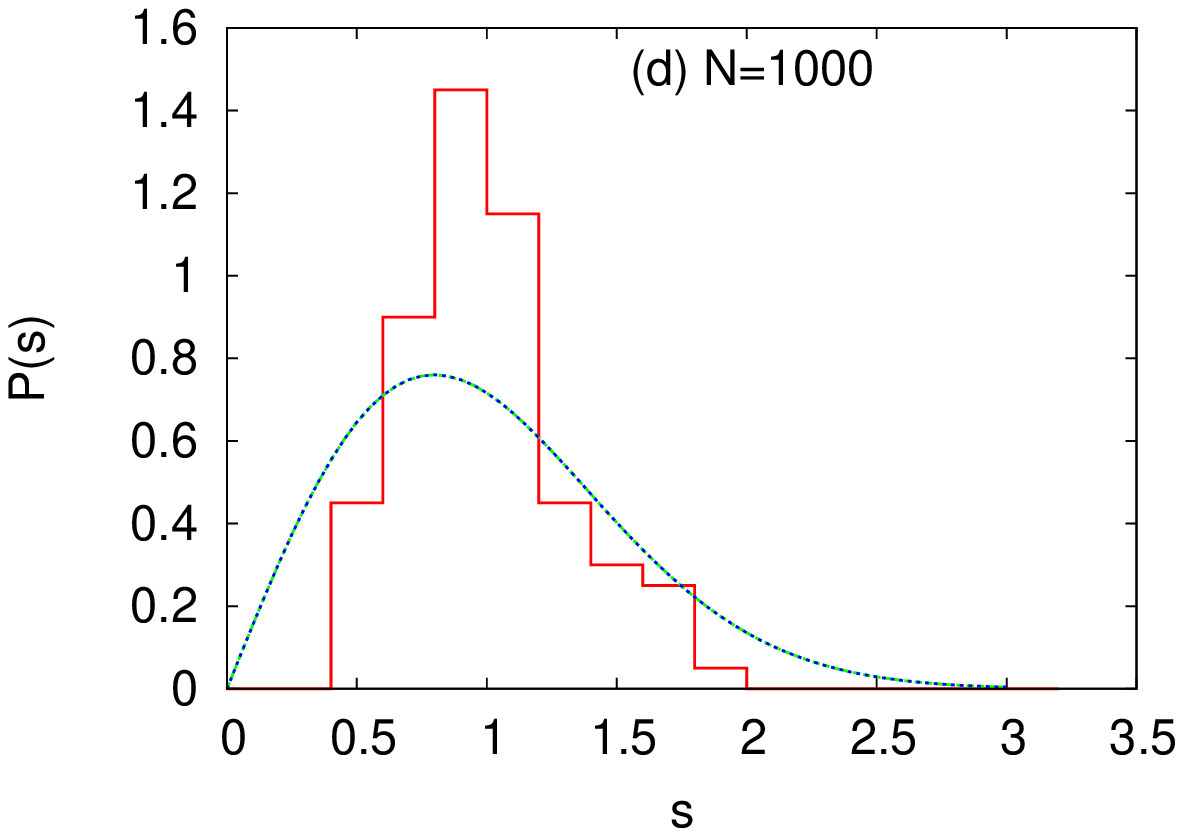}} \\
  
     \resizebox{40mm}{!}{\includegraphics[angle=0]{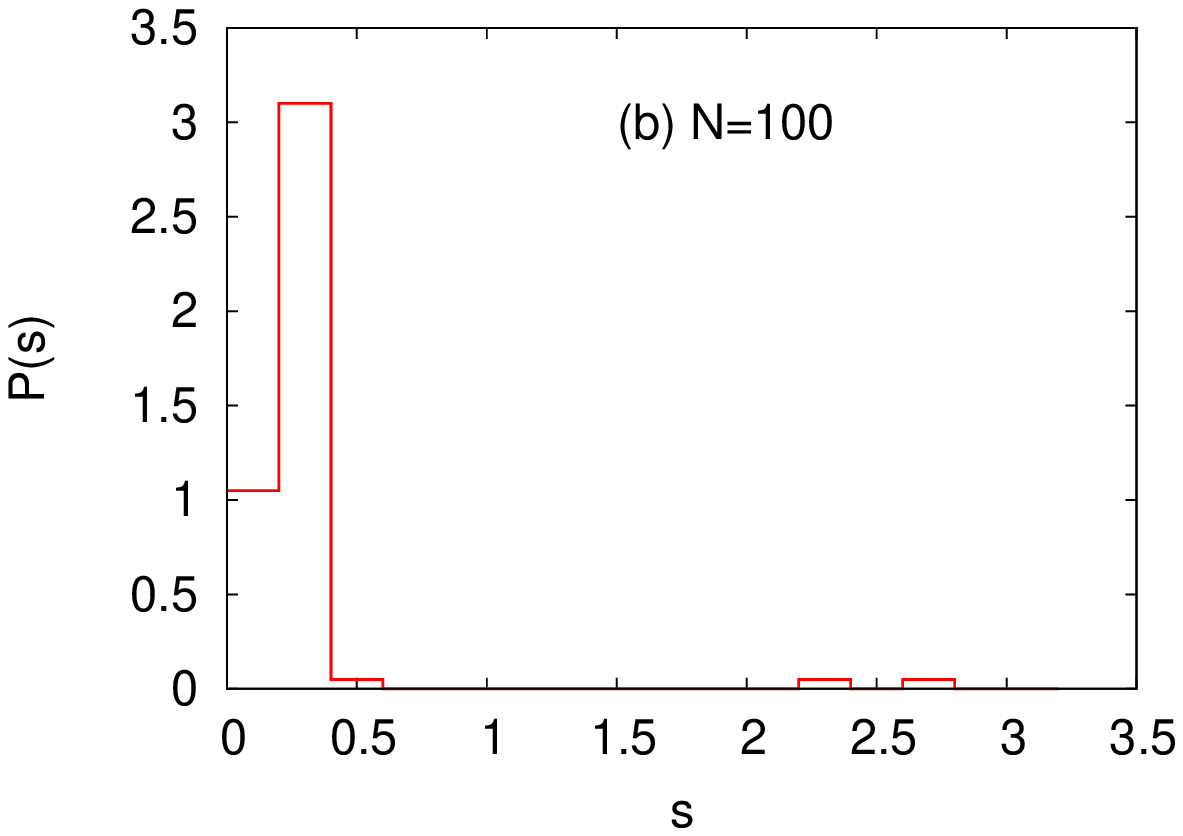}} &
      \resizebox{40mm}{!}{\includegraphics[angle=0]{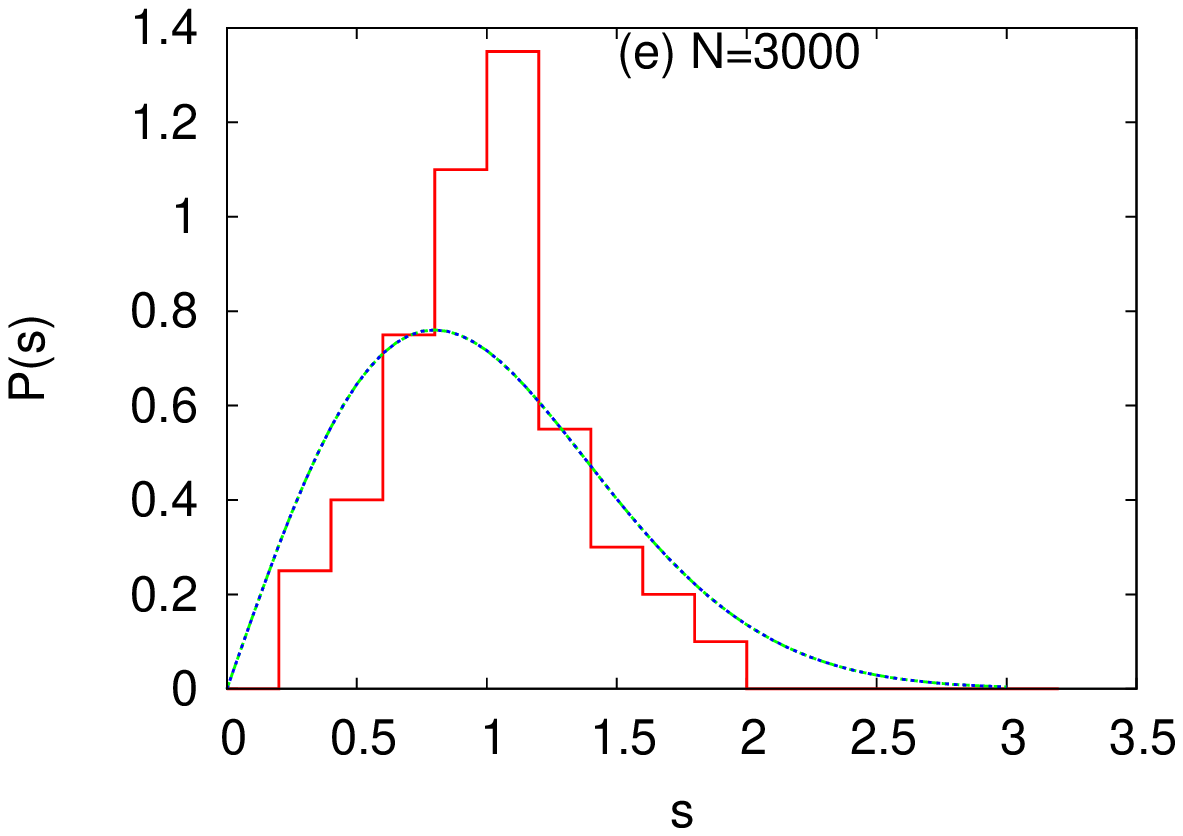}} \\

      \resizebox{40mm}{!}{\includegraphics[angle=0]{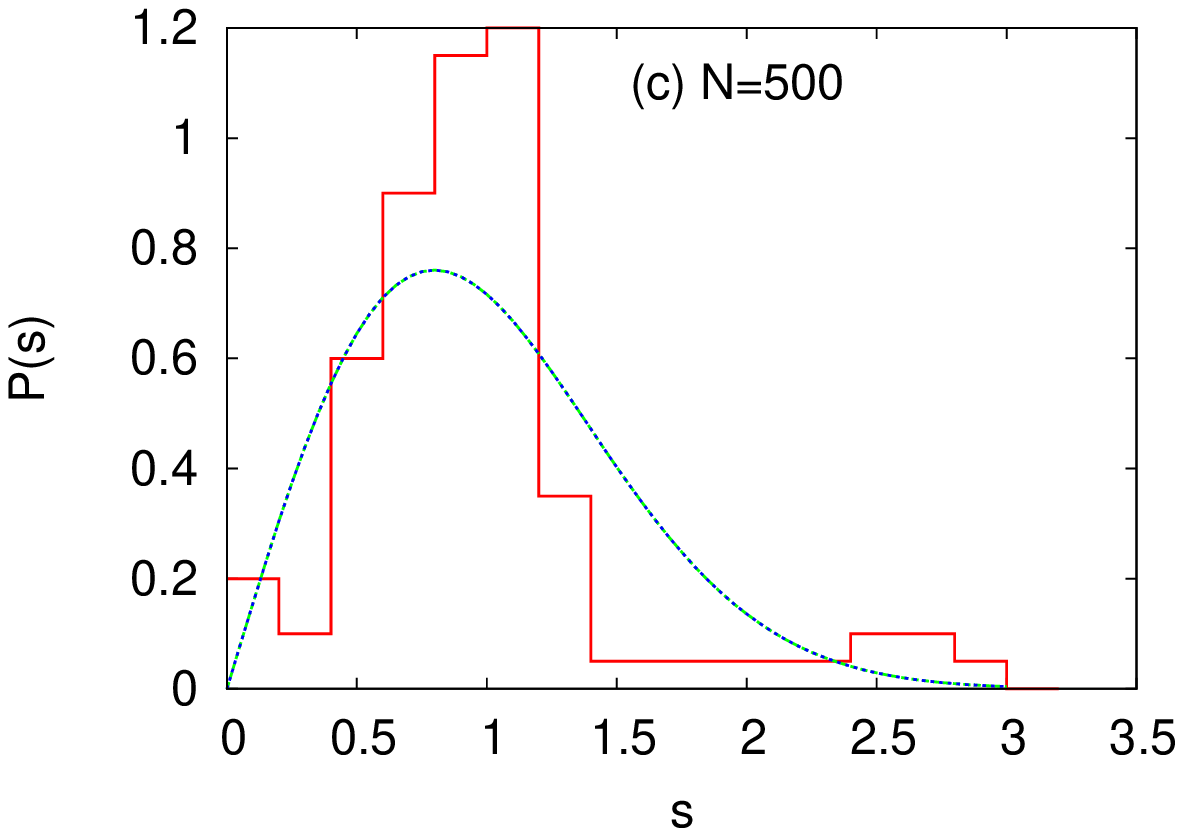}} &
      \resizebox{40mm}{!}{\includegraphics[angle=0]{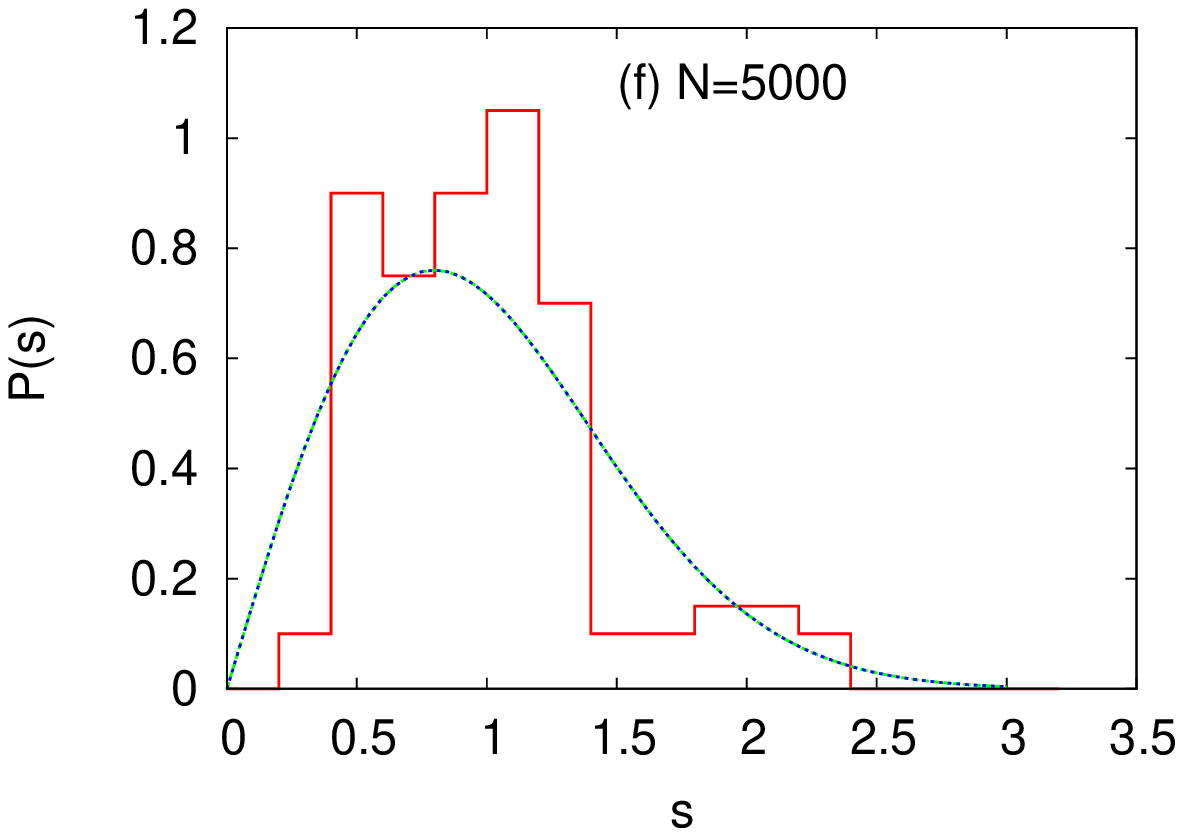}} \\
    \end{tabular}
  \end{center}
\caption{(color online) Histogram plot of $P(s)$ distribution {\it vs} $s$ for lowest 100 levels with different number of bosons in the trap. Blue lines are Wigner distribution.}
\end{figure}


\begin{figure}
  \begin{center}
    \begin{tabular}{cc}
      
      \resizebox{100mm}{!}{\includegraphics[angle=0]{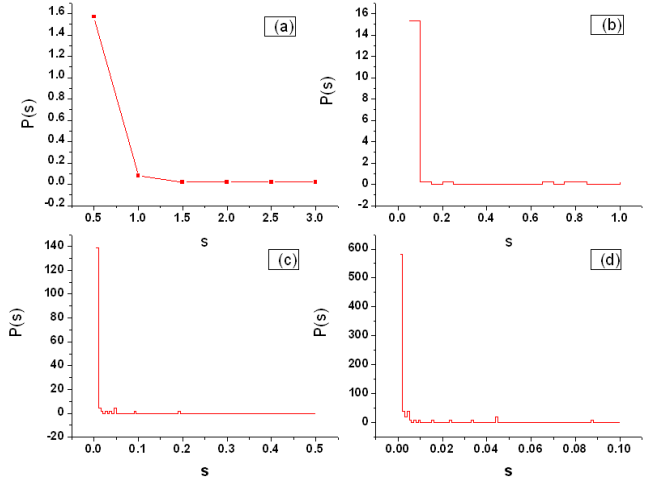}} \\
	 
    \end{tabular}
  \end{center}
\caption[angle=90]{(color online) Level spacing distribution $P(s)$ for the lowest 100 levels with 3 atoms in the trap. Fig.~4 (a)-(d) shows the Shnirelman peak in finer details.}
\end{figure}

\begin{figure}
  \begin{center}
    \begin{tabular}{cc}  
    \resizebox{70mm}{!}{\includegraphics[angle=0]{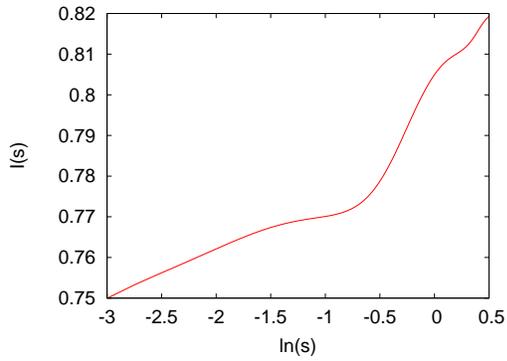}} \\	 
    \end{tabular}
  \end{center}
\caption[angle=0]{(color online) Plot of integral level spacing distribution $I(s)$ againt $ln (s)$}
\end{figure}

\begin{figure}
  \begin{center}
    \begin{tabular}{cc}
      \resizebox{40mm}{!}{\includegraphics[angle=0]{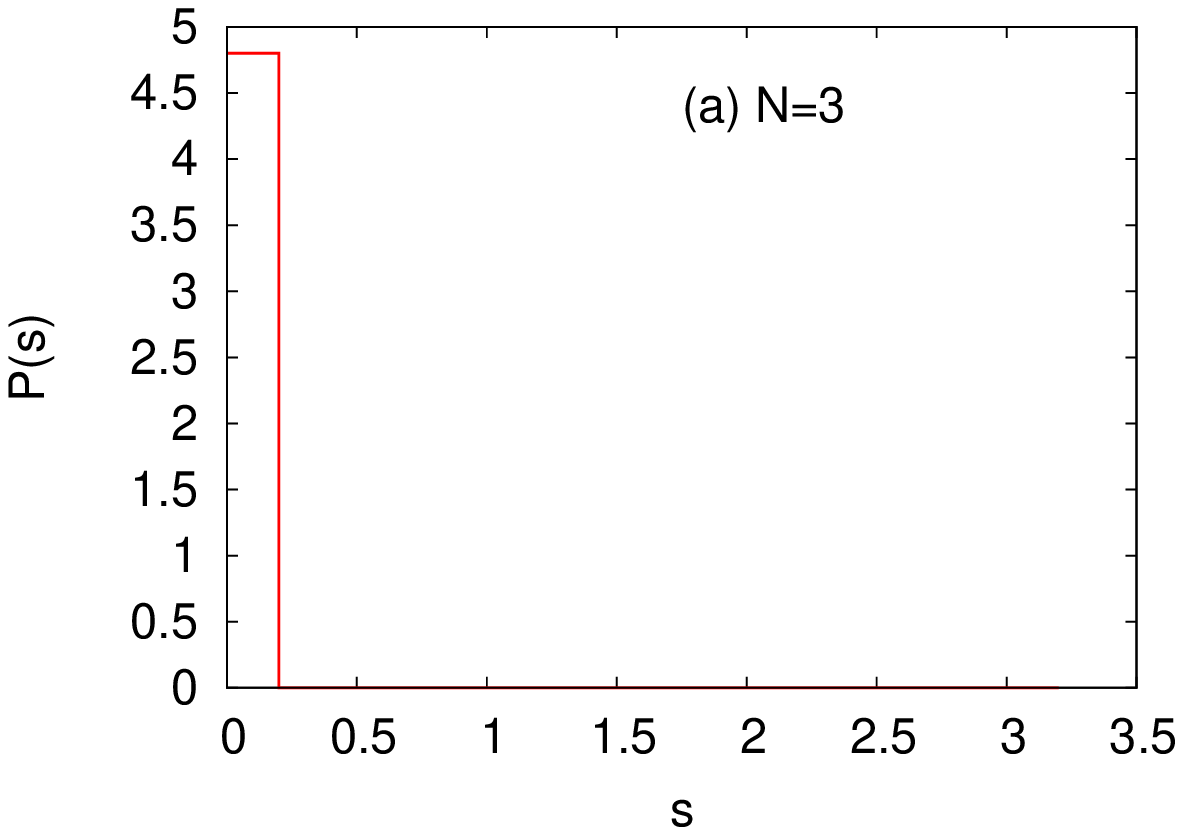}} &
       \resizebox{40mm}{!}{\includegraphics[angle=0]{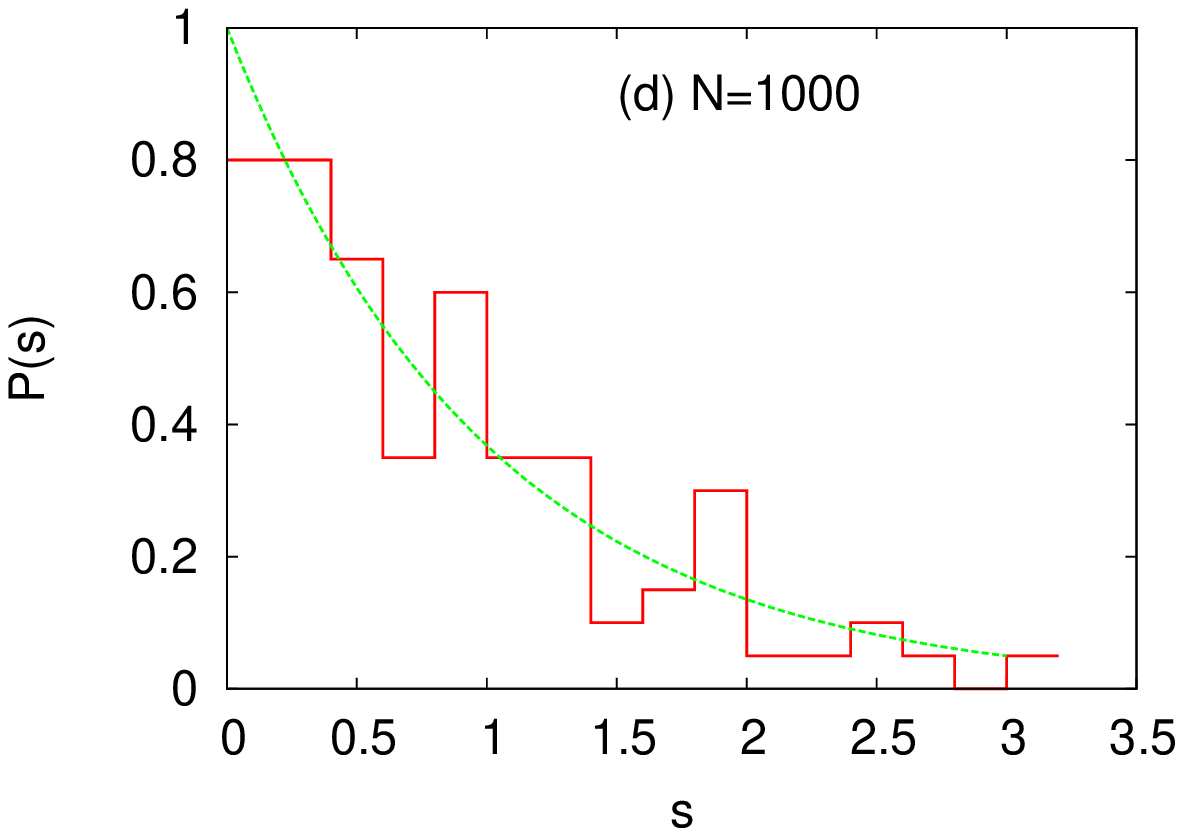}} \\
      \resizebox{40mm}{!}{\includegraphics[angle=0]{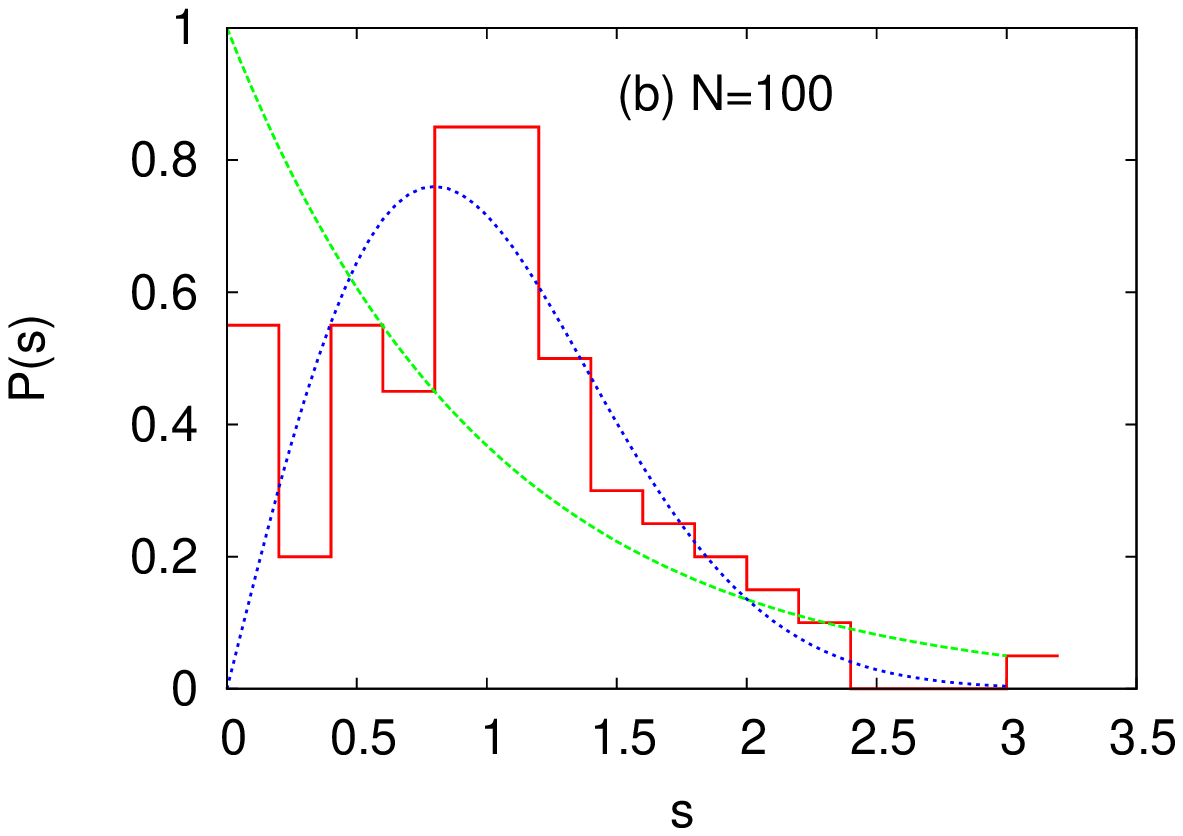}} &
      \resizebox{40mm}{!}{\includegraphics[angle=0]{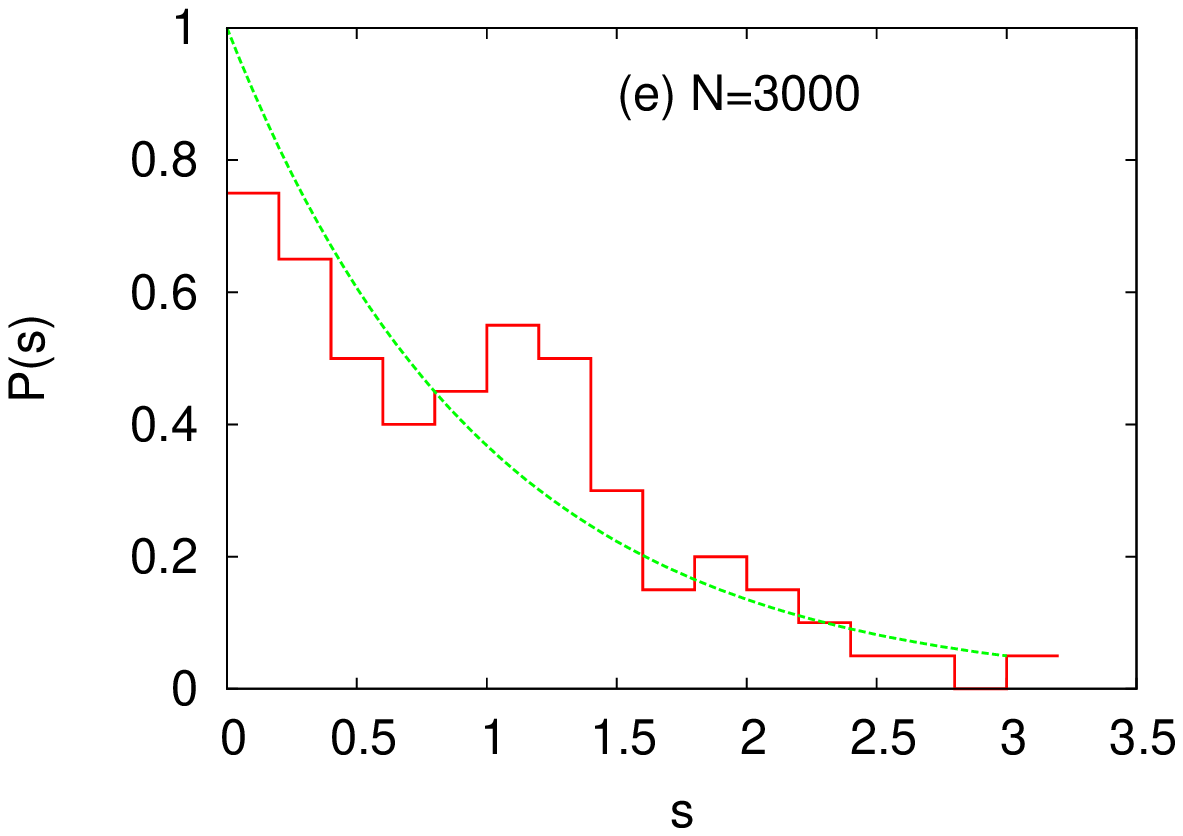}} \\

      \resizebox{40mm}{!}{\includegraphics[angle=0]{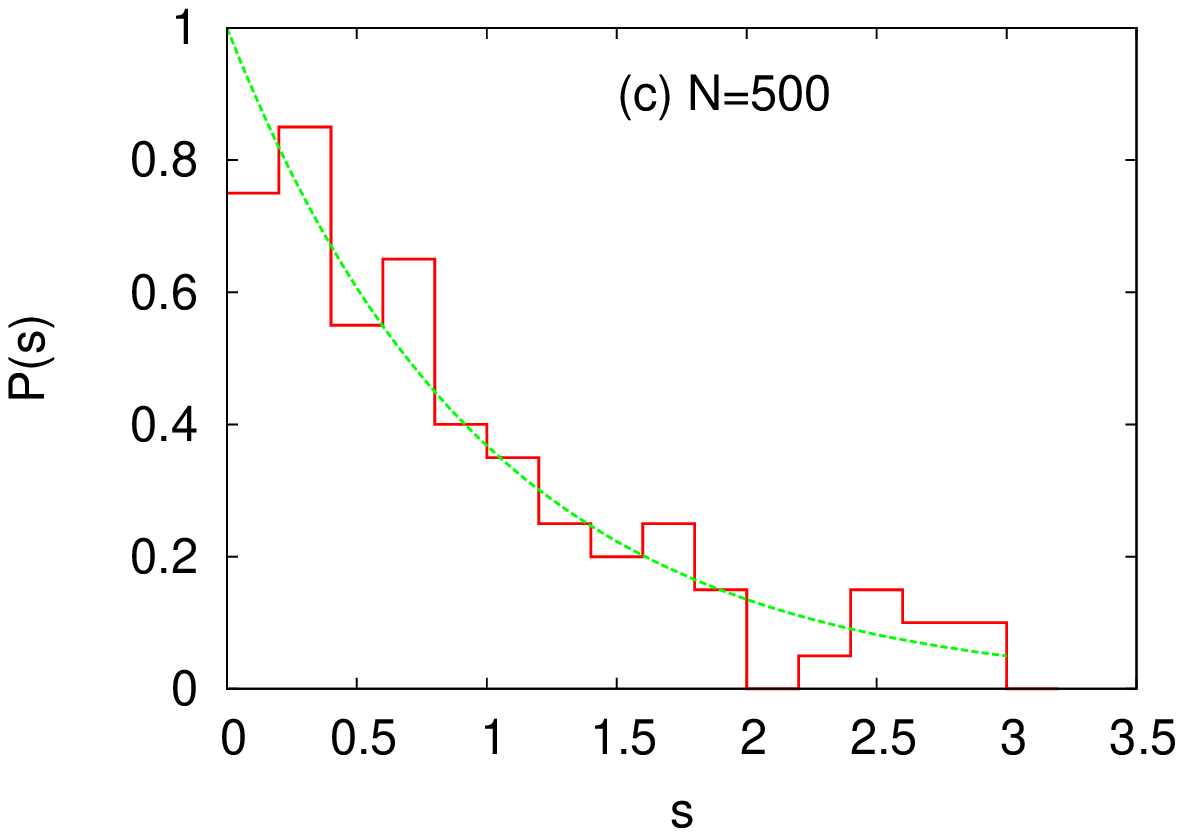}} &
      \resizebox{40mm}{!}{\includegraphics[angle=0]{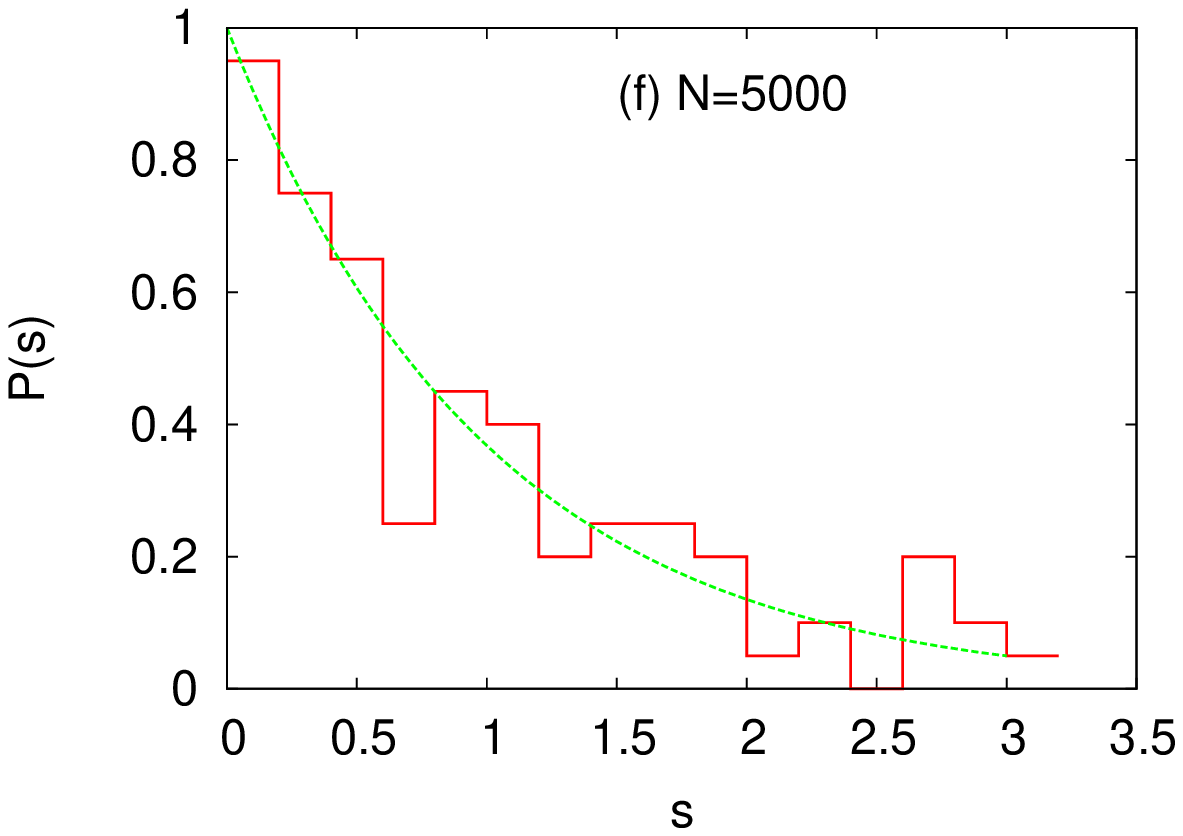}} \\
    \end{tabular}
  \end{center}
\caption{(color online) Histogram plot of $P(s)$ distribution {\it vs} $s$ for different the number of bosons ($N$) for energy-levels 3900 to 4000.  Green lines are Poisson distribution and Blue lines are Wigner distribution.}
\end{figure}
\vskip 0.5cm

\begin{figure}
  \begin{center}
    \begin{tabular}{cc}
       \resizebox{40mm}{!}{\includegraphics[angle=0]{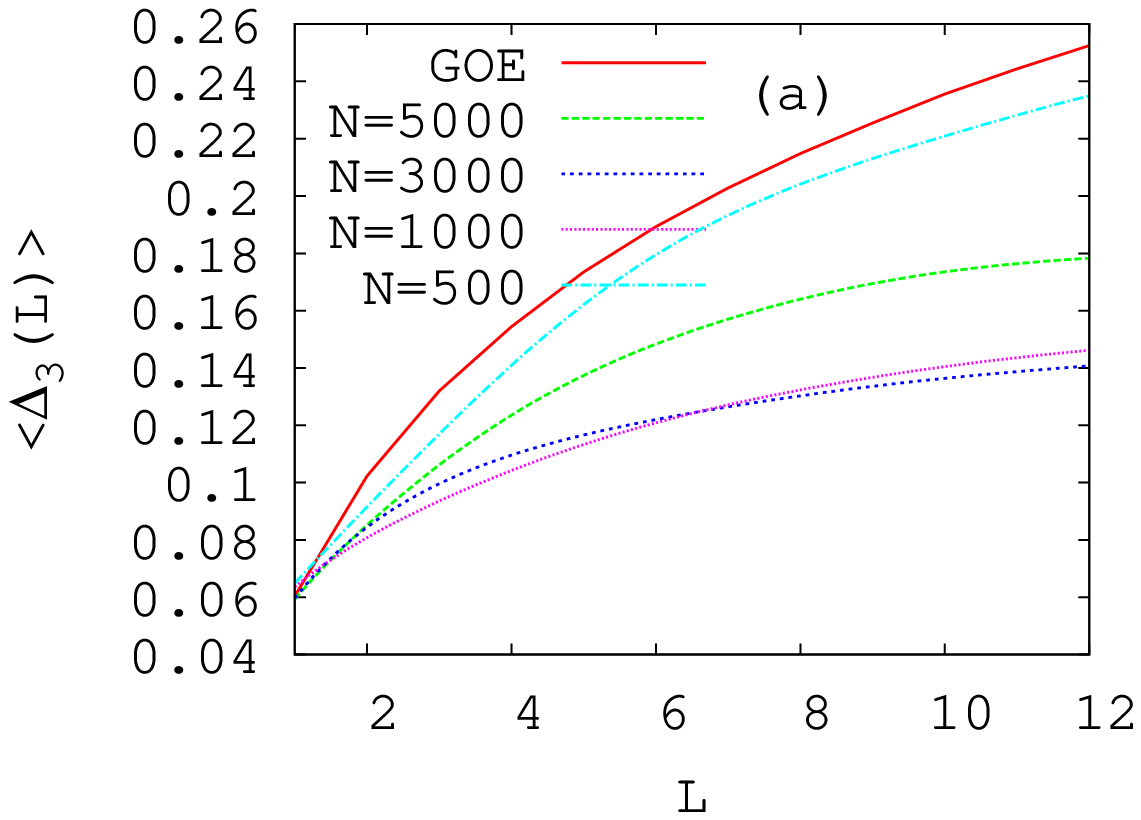}} &
       \resizebox{40mm}{!}{\includegraphics[angle=0]{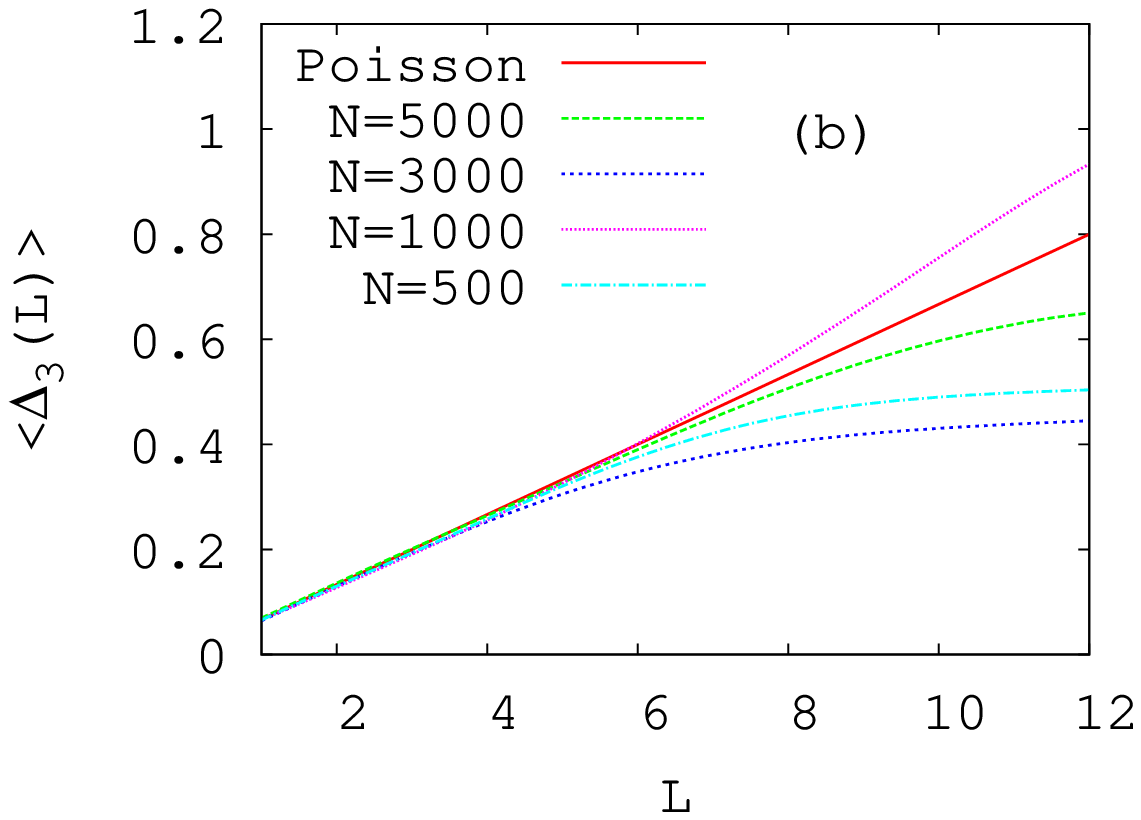}} \\  
\end{tabular}
  \end{center}
\caption {(color online) 
Spectral average $<\Delta_{3}(L)>$ computed for the Hamiltonian (1) with different number of interacting 
bosons ($N$) in the external trap {\it vs} $L$, (a) for lowest 100 energy levels and (b) for energy levels between 3900 and 4000.}
\end{figure}

\end{document}